\documentclass[floatfix,reprint,amsmath,amssymb,prapplied,superscriptaddress,aps]{revtex4-2}

\usepackage[dvipsnames]{xcolor}
\usepackage{graphicx}
\usepackage{amsmath,amssymb}
\usepackage{dcolumn}
\usepackage[utf8]{inputenc}
\usepackage[T1]{fontenc}
\usepackage{bm}
\usepackage{fontawesome5}
\usepackage{siunitx}
\usepackage{hyperref}

\hypersetup{
colorlinks=true, breaklinks=true, 
urlcolor=RoyalBlue, linkcolor=RoyalBlue, citecolor=RoyalBlue, 
pdftitle={},
pdfauthor={\textcopyright}, 
pdfsubject={},
pdfkeywords={}, 
pdfcreator={pdfLaTeX},
pdfproducer={LaTeX}
}

\begin{document}

\title{Micromagnetic structure of oxidized magnetite nanoparticles: sharp structural versus diffuse magnetic interface}

%\title{Micromagnetic structure of oxidized magnetite nanoparticles: diffuse magnetic interface at sharp structural interface}

\author{Elizabeth M. Jefremovas}
\email{elizabeth.jefremovas@.uni.lu}
\affiliation{Department of Physics and Materials Science, University of Luxembourg, 162A Avenue de la Faiencerie, L-1511 Luxembourg, Grand Duchy of Luxembourg}
\affiliation{Institute for Advanced Studies, University of Luxembourg, Campus Belval, L-4365 Esch-sur-Alzette, Luxembourg}
 \author{Michael P. Adams}
\affiliation{Department of Physics and Materials Science, University of Luxembourg, 162A Avenue de la Faiencerie, L-1511 Luxembourg, Grand Duchy of Luxembourg}
\author{Luc\'ia Gandarias}
\affiliation{Aix-Marseille Universit{\'e}, CEA, CNRS, BIAM, 13115 Saint Paul lez Durance, France}
\author{Lourdes Marcano}
\affiliation{Center for Cooperative Research in Biomaterials (CIC biomaGUNE), Basque Research and Technology Alliance, Donostia-San Sebasti\'an 20014, Spain}
\affiliation{Physics Department, University of Oviedo, C/ Calvo Sotelo, s/n. 33007 Oviedo, Spain}
\author{Javier Alonso}
\affiliation{Department CITIMAC, Faculty of Sciences, University of Cantabria, 39005 Santander, Spain}
\author{Andreas Michels}
\affiliation{Department of Physics and Materials Science, University of Luxembourg, 162A Avenue de la Faiencerie, L-1511 Luxembourg, Grand Duchy of Luxembourg}
\author{Jonathan Leliaert}
\affiliation{DyNaMat, Department of Solid State Sciences, Ghent University, 9000 Ghent, Belgium}

%%%%%%%%%%%%%%%%%%%%%%%%%%%%%%%%%%%%%%%%%%%%%%%%%%%
\begin{abstract}
The oxidation of magnetite to maghemite is a naturally occurring process that leads to the degradation of the magnetic properties of magnetite nanoparticles. Despite being systematically observed with traditional macroscopic magnetization measurement techniques, a detailed understanding of this process at the microscale is still missing. In this study, we track the evolution of the magnetic structure of magnetite nanoparticles during their oxidation to maghemite through numerical micromagnetic simulations. To capture realistic interparticle effects, we incorporate dipolar interactions by modeling the nanoparticles arranged in chains. Our computational results are benchmarked against experimental data from magnetotactic bacteria, studied over a time scale of years. To resolve the magnetization at the interface between both oxide phases, we propose spin-polarized small-angle neutron scattering (SANS), an experimental technique capable of probing magnetization textures at nanometer length scales. By analyzing the pair-distance distribution function extracted from SANS, we identify distinct signatures of magnetic disorder. Specifically, our findings suggest that the magnetization from the non-oxidized core region varies smoothly across the (structurally sharp) interface into the oxidized shell. The existence of such a diffuse magnetic interface may account for the superior magnetic properties of partially oxidized magnetite nanoparticles compared to fully converted maghemite samples.
\end{abstract}

%%%%%%%%%%%%%%%%%%%%%%%%%%%%%%%%%%%%%%%%%%%%%%%%%%%

\maketitle

%%%%%%%%%%%%%%%%%%%%%%%%%%%%%%%%%%%%%%%%%%%%%%%%%%%

\section{Introduction}

Magnetic nanoparticles (MNPs) are one of the leading agents of the new medicine era~\cite{wu2019magnetic, etemadi2020magnetic, farzin2020magnetic, lapusan2024advancing, rezaei2024magnetic}. With highly tunable properties and particle sizes of the order of biological entities (e.g., proteins, nucleid acids, viruses), their sizable magnetic moment allows, for example, the external activation using alternating magnetic fields, in this way enabling the remote generation of localized heat. In a therapeutic approach known as magnetic hyperthermia (MHT) this heat is harnessed to selectively destroy or debilitate cancerous cells (see, e.g., Refs.~\cite{pankhurst2003applications, pankhurst2009progress, smith2015nanoscale, perigo2015fundamentals, ortega2013magnetic, rubia2021whither} and references therein).

During the last years extensive efforts have been devoted to the engineering of magnetite MNPs with maximized heat release \cite{castellanos2021milestone, singh2016synthesis, rodriguez2018probing, sharifi2017combined}. The knowledge gathered so far points to the arrangement of magnetite MNPs in a chain configuration, which can be induced remotely in cubic morphologies~\cite{morales2021time}, as one of the top-notch arrangement in terms of bio-compatibility and maximized heat release during MHT treatments~\cite{valdes2020modeling, serantes2014multiplying, gandia2019unlocking, jefremovas2021nanoflowers, poon2025cubic}. However, there is a crucial aspect preventing the direct translation of these improved MNP designs into clinical results, which is the inevitable oxidation of magnetite over the time scale between the MNP synthesis and the clinical treatment. This naturally-occuring process degrades the magnetic properties of the magnetite phase, with the ensuing deterioration of the heating efficiency~\cite{jefremovas2020investigating, schwaminger2017oxidation, gehring2012oxidized}. Moreover, this process, which occurs in ambient atmosphere on time scales of weeks or months~\cite{mickoleit2023long, kim2021slow, schwaminger2017oxidation}, is sped-up for MNPs that are located within the tumoral area owing to their acid conditions~\cite{calaf2018oxidative, parolini2009microenvironmental, gandarias2023intracellular}. This requires the on-the-flow adjustment of the treatment, with a possible re-injection of MNPs after only 6~days~\cite{gandarias2023intracellular}.

In our work, we micromagnetically investigate progressively oxidized magnetite cubic MNP chains to gain insight in the concurrent magnetic degradation of the properties. Starting from a purely magnetite composition, with a defined uniaxial anisotropy direction $K_{\mathrm{u}}$, we model the effect of oxidation by including a maghemite shell of variable thickness $\delta$, which comprises random anisotropy directions. To ensure relevance to experimental systems, we follow the oxidation process of bacterial magnetosomes, i.e., pure magnetite MNPs biomineralized by magnetotactic bacteria (MTB)~\cite{faivre2008magnetotactic, fischer2011structural, marcano2022magnetic, fdez2013magnetite, gandia2022tuning}, keeping them intracellular over a time scale of years, using the measured values of the magnetic materials parameters as input for the simulations. Based on the numerical results for the spin structure, we compute the small-angle neutron scattering (SANS) cross section and the related pair-distance distribution function. This allows us to to identify the characteristic signatures of magnetic disorder and to resolve the magnetic correlations at the core@shell interface~\cite{zakutna2020field, michels2021magnetic, honecker2022using, lak2021embracing, gerina2024exploring}.

The article is organized as follows: In Sec.~\ref{exp}, we introduce the materials and methods used in this study. Section~\ref{results} discusses the results, with Sec.~\ref{experiments} presenting the experimental results that are used as input for the material parameters in the micromagnetic simulations. In Sec.~\ref{micromagnetic}, we discuss the results of the numerical micromagnetic calculations, which are used to model the progressive formation of the maghemite shell. This is done by calculating the coercivity and by comparing the results for different layer thicknesses $\delta$. In Sec.~\ref{SANS_pr}, we show the neutron scattering results for the pair-distance distribution function $p(r)$. We consider that $p(r)$ is an important quantity where the most pronounced effect of the interface-thickness variation becomes visible and that can also be easily detected experimentally. Finally, Sec.~\ref{conclusions} summarizes the main findings of this work. The Supplementary Information~\cite{SM_Nanoscale_2025} provides further details on the structural and magnetic characterization of the magnetite MNPs and the main expressions for the SANS observables.

%%%%%%%%%%%%%%%%%%%%%%%%%%%%%%%%%%%%%%%%%%%%%%%%%%%

\section{Materials and methods}
\label{exp}

Magnetotactic bacteria \textit{Magnetospirillum gryphiswaldense} (MSR-1) strain MSR-1 (DMSZ 6631) and \textit{Magnetospirillum magneticum} (AMB-1) strain AMB-1 have been cultured microaerobically at $28^{\circ} \, \mathrm{C}$ in flask standard medium, as described by Heyen and Sch\"uler~\cite{heyen2003growth}, and supplemented by $100 \, \mu \mathrm{M}$ iron(III)-citrate to support magnetosome formation. Briefly, the culture was carried out in $1 \, \mathrm{L}$~bottles at $28^{\circ} \, \mathrm{C}$ under microaerobic conditions (bottles loosely closed and without shaking for MSR-1; and bottles filled to the top and with the caps completely closed for AMB-1). Cells were collected after $96 \, \mathrm{h}$ when well-formed magnetosomes were present \cite{gandia2022tuning}. The cells were harvested by centrifugation ($8000 \, \mathrm{g}$, $15$~minutes, 4\textsuperscript{$\circ$}C) and the pellets were resuspended in PBS and fixed with 2\% glutaraldehide overnight. Then, the bacteria were washed 3 times with PBS and 2 times with filtered MilliQ water. Lastly, the bacterial pellets were freeze-dryed, resulting in a powder sample. The samples were kept under vacuum (base pressure $p \lesssim 8.8 \times 10^{-2} \, \mathrm{mbar}$) for the whole duration of the study.

DC magnetization measurements were performed using a Quantum Design QD-MPMS SQUID magnetometer in the temperature range between $5$$-$$300 \, \mathrm{K}$. The samples consist of magnotosome chains kept within the bacteria body, each of the bacteria randomly oriented with respect to each other. $M(T)$~curves were measured from $10$ to $300 \, \mathrm{K}$, following a zero-field-cooling/field-cooling protocol (ZFC-FC): the samples were cooled in the absence of an applied field from $300 \, \mathrm{K}$ to $5 \, \mathrm{K}$. At $5 \, \mathrm{K}$, a fixed magnetic field of $5 \, \mathrm{mT}$ was applied and the magnetization was measured upon warming to $300 \, \mathrm{K}$ (ZFC). With the field still on, the sample was cooled to $5 \, \mathrm{K}$ and the magnetization was measured upon warming to $300 \, \mathrm{K}$ (FC). $M(H)$~loops were measured at $300 \, \mathrm{K}$ and $5 \, \mathrm{K}$ for applied fields up to $2 \, \mathrm{T}$.

%The high-sensitivity of the SQUID ($\sim 10^{-7} \, \mathrm{emu}$) allowed us to detect the small amounts of the MNPs ($m \approx 0.9 \, \mathrm{mg}$ of freeze-dried bacteria) \textcolor{red}{(why is this important? just wanted to make the point that the amount of magnetite per bacteria is very low, but you are right, let's remove)}.

Micromagnetic simulations were performed using the software package MuMax3 (version 3.11)~\cite{vansteenkiste2014design}. The isotropic exchange interaction, an uniaxial magnetic anisotropy, the Zeeman energy as well as the magnetodipolar interaction have been taken into account. The following materials parameters were used:~For the Fe\textsubscript{3}O\textsubscript{4} phase, the saturation magnetization has been fixed to $M_{\mathrm{s}} = 0.486 \, \mathrm{MA/m}$, the exchange-stiffness constant to $A = 13 \, \mathrm{pJ/m}$, and the uniaxial anisotropy constant was taken as $K_{\mathrm{u}} = 35 \, \mathrm{kJ/m^3}$, based on both experimental and literature data~\cite{gandia2024exploring, witt2005three, jefremovas2022modifying}. For the maghemite phase, a reduced $M_{\mathrm{s}}$ of $0.412 \, \mathrm{MA/m}$ has been assumed keeping the uniaxial anisotropy at its bulk value. The anisotropy direction of the magnetite phase has been fixed to follow the chain axis, which is oriented at angles between $12$$-$$20^{\circ}$ relative to the crystallographic $\left< 111 \right>$ direction~\cite{orue2018configuration}. In the present simulations, this angle has been fixed to $15^{\circ}$. To simulate the magnetically-disordered environment of the maghemite phase, the surface region has been subdivided into grains using a Voronoi tesselation~\cite{Lel2014}. The shell then consists of a collection of grains, where the uniaxial anisotropy points into random directions. The exchange coupling at the intergrain boundaries has been reduced by a factor $0 < k \leq 1$ to model the effect of disorder:~A value of $0.75 \times A$ ($k=0.75$) reproduces the experimental results well. The grain size $D$ has been slightly varied to follow the shell thickness $\delta$; specifically, we used $D = 2 \, \mathrm{nm}$ for $\delta = 2 \, \mathrm{nm}$, $D = 4 \, \mathrm{nm}$ for $4 \, \mathrm{nm}$ and $8 \, \mathrm{nm}$ shells, and $D = 5 \, \mathrm{nm}$ for $\delta = 5 \, \mathrm{nm}$ and $10 \, \mathrm{nm}$. The sample volume was discretized into cells with a typical size of $2 \, \mathrm{nm}$. This cell value is well below the exchange length $l_\mathrm{ex} = \sqrt{2A/(\mu_0 M_{\mathrm{s}}^{2})} \cong 9.4 \, \mathrm{nm}$, ensuring that nanometer-scale spatial variations of the magnetization can be resolved. The energy minimum was obtained by using the steepest descent method~\cite{Exl2014}, encoded in the ``minimize'' function of MuMax3. To reproduce the random orientations of the MNPs in the experiments, the external magnetic field $\vec{H}$ has been rotated from $0$ to $90^{\circ}$, and we have repeated the simulations for $10$ different random seeds to obtain statistically significant results.

%(\textcolor{red}{only ???. We typically use $\sim$ 100 averages. In this case, what we are varying is the randomness of the surface grains. It is already good with 5 different realizations, but 10 is excellent. I discussed it with Jonathan and he approved. I think for SANS is a little bit different, since the calculations you run are a little bit faster (in general, or at least, up to my experience). Take into account that I am relaxing/minimizing the structure at each field step, so in total, every loop includes a super large number of steps. The results we obtain deliver a very good compromise between accuracy and computing time. running 100 seeds will not provide greater accuracy than running 10 but with finer field steps}).

The calculation of the polarized small-angle neutron scattering (SANS) observables follows the procedure detailed in~\cite{michels2021magnetic, vivas2020toward, pratami2023micromagnetic, adams2024framework, sinaga2024neutron}. Simulations were performed on a single, oriented cuboidal particle. Briefly, the resulting discrete real-space magnetization vector of the MNP, $\mathbf{M}_i(\mathbf{r}_i)$, where the subscript ``$i$'' refers to the $i$th cell, is Fourier transformed to obtain the magnetic SANS cross section for the particular scattering geometry, where the applied magnetic field is perpendicular to the incoming neutron beam. The magnetic scattering cross section on the two-dimensional detector is then averaged to obtain the one-dimensional SANS intensity profile $I_{\mathrm{sf}}(q)$ from which the pair-distance distribution function $p(r)$ is numerically computed via Fourier transformation. In this paper, we have computed $p(r)$ results for the purely magnetic spin-flip SANS cross section, which can be obtained from uniaxial neutron polarization analysis experiments (e.g., \cite{michels2010epjb,kryckaprl2014,zakutna2020,malyeyev2022,ukleev2024}) and does not contain the nonmagnetic (structural) scattering contribution. See the Supplementary Information~\cite{SM_Nanoscale_2025} for further details.

%%%%%%%%%%%%%%%%%%%%%%%%%%%%%%%%%%%%%%%%%%%%%%%%%%%%%%%%%%%%%%%%%%%%%%%%%%%

\section{Results and discussion}
\label{results}

%%%%%%%%%%%%%%%%%%%%%%%%%%%%%%%%%%%%%%%%%%%%%%%%%%%%%%%%%%%%%%%%%%%%%%%%%%%

\subsection{Magnetic properties of oxidized bacterial magnetosomes}\label{experiments}

We have followed the oxidation process of bacterial magnetosomes kept intracellular at a time scale of years, measuring the degradation of their magnetic properties. These values, together with the morphology of the bacteria at $t = 0$ (shown in the Supplementary Information Figures~1 and 2) are used as input material parameters to our simulations, connecting our calculations with experimental values. 

\begin{figure}
\centering
\resizebox{1.0\columnwidth}{!}{\includegraphics{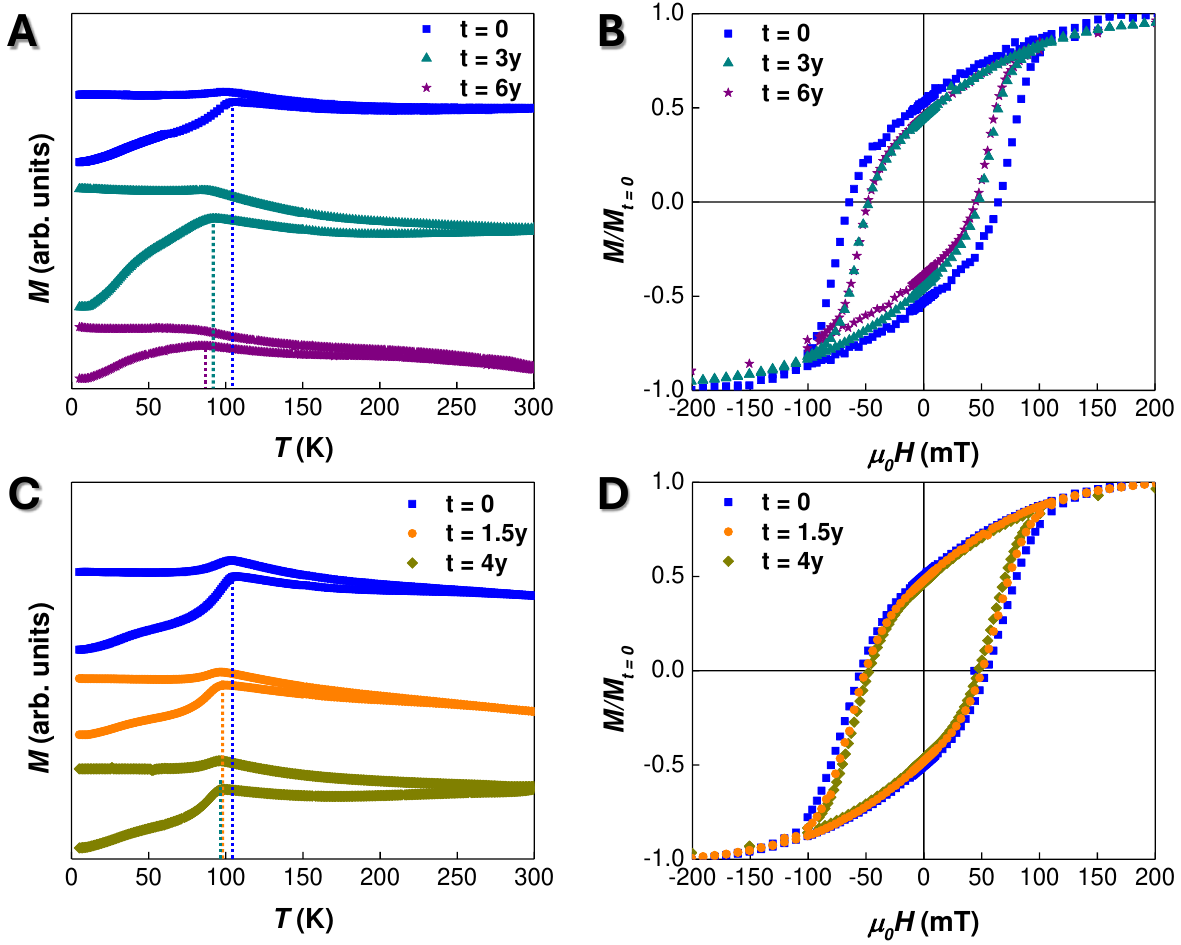}}
    \caption{(\textbf{A}) and (\textbf{C}): Zero-field-cooling/field-cooling (ZFC-FC) $M(T)$ curves for MSR-1 and AMB-1, respectively, measured at 0, 1.5, 3, 4, and 6~years. The position of the Verwey transition is marked by dotted vertical lines. (\textbf{B}) and (\textbf{D}): Hysteresis loops $M(H)$ measured at $T = 5 \, \mathrm{K}$ normalized by the value of $M(t=0)$ for each bacterial strain.}
    \label{experimental_magnetization}
\end{figure}

In order to detect the variations of the magnetic properties with time, we have studied the same batch of MSR-1 and AMB-1 bacteria, starting at $t=0$ (fresh bacteria) along a long period of time (3 and 6~years for MSR-1 and 1.5 and 4~years for AMB-1). Unlike the reported studies with uncapped inorganically-synthesized magnetite nanoparticles~\cite{kim2021slow, schwaminger2017oxidation}, the proteolipidic membrane of the magnetosomes, in addition to the bacterial body, acts as a strong barrier against oxidation, in this way slowing down the oxidation process from a few months to years.

Figure~\ref{experimental_magnetization} features the magnetization data of MSR-1 and AMB-1 bacteria. Both $M(T)$ ZFC-FC curves exhibit a pronounced irreversibility throughout the entire temperature range, with a blocking temperature being above 300 K, consistent with the large size of the magnetosomes. Similarly, the ZFC curve of the two studied strains present a sudden drop in the magnetization at $T_{\mathrm{V}}$, a typical signature of the Verwey transition, which ensures the presence of magnetite in the bacteria~\cite{walz2002verwey, yu2014verwey}. At $t = 0$, the $T_{\mathrm{V}}$ is found at $104 \pm 2 \, \mathrm{K}$ ($106 \pm 2 \, \mathrm{K}$ for AMB-1), a value that is lower than the bulk value of magnetite ($T_{\mathrm{V}} = 120 \, \mathrm{K}$ \cite{walz2002verwey}), yet typically reported for magnetotactic bacteria~\cite{marcano2017influence, gandia2024exploring, jefremovas2022modifying, gandia2022tuning, gandia2020elucidating}. The decrease with respect to the bulk value is usually ascribed to microstrain disorder, yielding slight distortions from the perfect bulk magnetite crystalline structure, which is supported by the nonzero microstrain value shown in our XRD analysis (see the Supplementary Information~\cite{SM_Nanoscale_2025}). After 1.5 years, this transition decreases by $7 \, \mathrm{K}$ in AMB-1, remaining at $T_{\mathrm{V}} = 98 \, \mathrm{K}$ after 4~years, thus, yielding a total reduction of $\approx 7.5 \, \%$ after 4~years. This reduction is of the same order of magnitude as the one experienced by \textit{M. griphyswaldense} after 3~years. In this case, the $T_{\mathrm{V}}$ decreases to $T_{\mathrm{V}} = 95 \pm 2 \, \mathrm{K}$, i.e., $\approx 8.6 \, \%$. The reduction of $T_{\mathrm{V}}$ is an indication of an oxidation process of magnetite to maghemite, since the latter does not undergo the Verwey transition on behalf of its lack of Fe\textsuperscript{2+} ions~\cite{laurent2010magnetic, faivre2016iron, millan2007surface}. Further increase of time results in a further reduction of $T_{V}$, as expected from the reduced energy scale of the transition in the presence of the increased disorder due to the oxide shell. In this sense, MSR-1 reduces $T_{\mathrm{V}}$ to $88 \, \mathrm{K}$ after 6~years, corresponding to an $\approx 15 \, \%$ reduction of the initial value. It is interesting to note that the decrease in $T_{\mathrm{V}}$ obtained in both samples is appreciably smaller than the one typically observed for synthetic magnetite MNPs; for example, Kim et al.~\cite{kim2021slow} reported a $\approx 25 \, \%$ decrease of $T_{\mathrm{V}}$ in 44 nm magnetite MNPs after 3 years. This implies that, by preserving the magnetosomes inside the bacterial body, we successfully manage to slow down the oxidation process that naturally happens in magnetite.

The difference in the shape of the Verwey transition observed in the ZFC-FC curves of both bacterial strains also provides insights into the bacterial chain configuration. Unlike bulk magnetite, where the Verwey transition is sharp and abrupt, in magnetotactic bacteria, the magnetite MNPs are synthesized by living entities through a complex biomineralization process, which comprises a compendium of cellular processes involving a large set of specialized genes~\cite{muela2016optimal, munoz2020magnetosomes, wan2024biomineralization, amor2020iron, nudelman2018understanding, ben2021current}. As a result, inhomogeneities in the magnetosome shape and size occur, yielding a broader transition in the ZFC-FC curves compared to bulk magnetite~\cite{walz2002verwey}. Note particularly the case of MSR-1 in Fig.~\ref{experimental_magnetization}(\textbf{A}). As introduced in the previous section, the biomineralization process of MSR-1 and AMB-1 are different, as the former tends to grow magnetosomes arranged in one single long chain, whereas the latter tends to nucleate magnetosomes at different positions, forming smaller chains~\cite{fdez2013magnetite, fdez2020magnetotactic, fdez2021nature, muela2016optimal, gandia2022tuning, amor2018iron}. This results in a broader size distribution of the magnetosomes for MSR-1, and thus, in a broader Verwey transition compared to the one of AMB-1.

%The oxidation of the magnetite phase towards maghemite also modifies the hysteresis loops. It is worth mentioning that, at $T > T_{\mathrm{V}}$, the magnetocrystalline anisotropy is dominated by the cubic term $K_{\mathrm{c}}$ and $ \left< 111 \right>$ easy axes from magnetite. This means that as long as the presence of magnetite is significant in our magnetosomes the coercivity will be dominated by this phase, and the magnetic properties remain almost unaffected. As an example, we have included in the Supplementary Information~\cite{SM_Nanoscale_2025} the comparison between $t = 0$ and $t = 6 \, \mathrm{years}$ for MSR-1, corresponding to the highest oxidation. The coercivity of both hysteresis loops nearly overlap. This has a major advantage for hyperthermia applications, as the clinical temperatures $T \sim 320 \, \mathrm{K} \gg T_{\mathrm{V}}$, thus, leaving the magnetic properties almost unaffected for small oxidation rates (in our case, around $\approx$$15 \, \%$).
%uniaxial $K_{\mathrm{u}}$
To capture the consequences of the oxidation, we have analyzed the measurements performed at $T = 10 \, \mathrm{K} \ll T_{\mathrm{V}}$, where the anisotropy is dominated by the uniaxial $K_{u}$ term. Above $T_{\mathrm{V}}$, the magnetocrystalline anisotropy is dominated by the cubic term $K_{\mathrm{c}}$ with $ \left< 111 \right>$ easy axes, which is less sensitive to structural distortions or surface oxidation, especially when magnetite remains the dominant phase. We refer the reader to the Supplementary Information ~\cite{SM_Nanoscale_2025}, where the comparison between the hysteresis loops at $T = 300\; \mathrm{K}$ for MSR-1 measured at $t = 0$ and $t = 6 \, \mathrm{years}$ is provided. 

Figures~\ref{experimental_magnetization}(\textbf{B}) and (\textbf{D}) include the $M(H)$ hysteresis loops measured at $T = 10 \, \mathrm{K}$, for MSR-1 and AMB-1, respectively. At such a temperature, the anisotropy is dominated by the uniaxial term $K_{\mathrm{u}}$. A reduction of the coercive field, $H_{\mathrm{c}}$, can be detected, serving as an indication of the presence of a disordered surface at the magnetosomes~\cite{roca2007effect, fiorani2002magnetic}. In this way, for MSR-1, $H_{\mathrm{c}}$ reduces from $64\, \mathrm{mT}$ to $49 \, \mathrm{mT}$ after 3~years, i.e., almost by $25 \, \%$. Measurements carried out after 6~years indicate no significant further oxidation, and the $H_{\mathrm{c}}$ value stays almost constant ($48 \, \mathrm{mT}$). For AMB-1, $H_{\mathrm{c}}$ reduces from $54 \, \mathrm{mT}$ to $50 \, \mathrm{mT}$ after 1.5~years, and further to $47 \, \mathrm{mT}$ after 4~years, i.e., a reduction of almost $15 \, \%$ of the initial value. The presence of smaller magnetosomes in the chains of MSR-1, more sensitive to oxidation on behalf of their larger surface-to-volume ratio, could account for this difference.

%Reduction in Ms: In \cite{ge2014effects} they also report a strange behaviour of Ms.

%%%%%%%%%%%%%%%%%%%%%%%%%%%%%%%%%%%%%%%%%%%%%%%%%%%%%%%%%%%%%%%%%%%%%%%%%%%

\subsection{Micromagnetic simulations of magnetite@maghemite nanoparticles}
\label{micromagnetic}

\begin{figure*}
    \centering
    \resizebox{2.0\columnwidth}{!}{\includegraphics{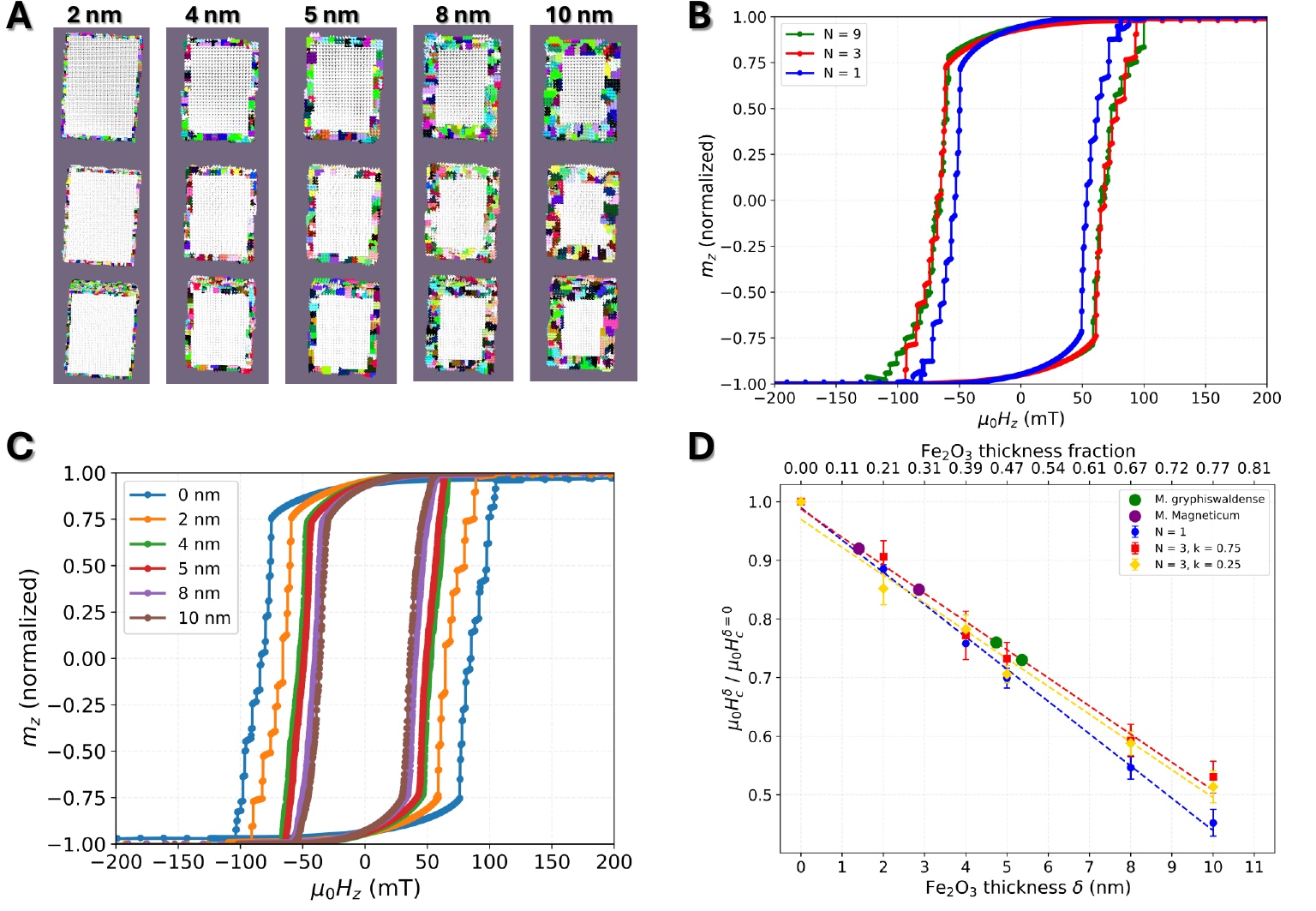}}
    \caption{(\textbf{A})~Microstructures of the simulated cuboid chains for different Fe\textsubscript{2}O\textsubscript{3} shell thicknesses $\delta$. The size of an individual particle is $48 \times 48 \times 64 \, \mathrm{nm}$, and panel~(\textbf{A}) shows chains of $N=3$ dipolarly-coupled particles. The color scheme represents the uniaxial anisotropy direction for all the magnetic cells. The ones within the magnetite core (white color) follow the $\left< 111 \right>$ direction rotated by $15^{\circ}$ with respect to the $z$ axis to mimic the chain configuration~\cite{orue2018configuration}. For the maghemite shell, the direction of $K_{\mathrm{u}}$ is random in each grain, represented by the different colors. (\textbf{B})~Hysteresis loops for $N = 1$, 3, and 9~chains. The coercivity $H_{\mathrm{c}}$ increases from $N = 1$ to $N > 1$ with no significant differences between $N = 3$ and $N = 9$. (\textbf{C})~Simulated hysteresis loops for a chain of $N = 3$ cuboids for different $\delta$. The coercive field $H_{\mathrm{c}}$ decreases as $\delta$ increases. (\textbf{D})~$H_{\mathrm{c}}$ versus layer thickness $\delta$: values are calculated for $N = 1$ (blue), $N = 3$ with an intergrain exchange-coupling reduction of $k = 0.25$ (yellow), and $k = 0.75$ (red) cuboid chains. For clarity, the $H_{\mathrm{c}}$ values have been normalized to the value at $\delta = 0 \, \mathrm{nm}$. $H_{\mathrm{c}}$ decreases linearly with $\delta$. We have included the experimental data for \textit{Magnetospirillum gryphiswaldense} and \textit{Magnetospirillum magneticum} to estimate the $\delta$ agreeing with the observed $H_{\mathrm{c}}$ reduction.}
    \label{simulations}
\end{figure*}

Our magnetic measurements provide indirect evidence for the oxidation of the magnetosomes. This finding is based on: (i)~a less abrupt and temperature-shifted Verwey transition and (ii)~a decreased coercivity at $T = 10 \, \mathrm{K}$. To model and quantify the oxidation process by determining the amount of magnetite that transforms into maghemite, we have numerically simulated the hysteresis loops of individual magnetosomes and the chain arrangements. Our simulations approximate the magnetosomes as cuboids of dimensions $44 \times 44 \times 64 \, \mathrm{nm}^3$. This geometry is also similar to the elongated magnetosomes of \textit{M. blakemorei} strain~\cite{bazylinski1988anaerobic, silva2013optimization, kalirai2013anomalous}, and is representative for the magnetosomes studied in this work (see the Supplementary Information~\cite{SM_Nanoscale_2025} for TEM images). To account for the oxidation, we have simulated a core@shell structure, consisting of a large magnetite core with a defined uniaxial anisotropy direction that is surrounded by a polycrystalline surface layer of maghemite. We have included a sketch of our simulated cuboid chains in Fig.~\ref{simulations}(\textbf{A}), which also reproduces the slight rotation of the magnetosomes ($15^{\circ}$)~\cite{orue2018configuration} to make our model as close to reality as possible.

Figure~\ref{simulations} (\textbf{B}) features the hysteresis loops for $N = 1$, $3$, and $9$ cuboids arranged in chains for a pure magnetite phase (i.e., no oxidation). We have performed our simulations at $T = 0 \, \mathrm{K}$ to remove the thermal noise and to closely mimic the $T = 5 \, \mathrm{K}$ experimental conditions. First, we observe a significant increase of the coercivity (by about $20 \, \%$) when $N > 1$. This difference is triggered by the dipolar interaction between the cuboids. Comparing the $N = 3$ and $N = 9$ data, no significant differences are found ($\mu_0 H_{\mathrm{c}} = 68 \, \mathrm{mT}$ and $69 \, \mathrm{mT}$, respectively). This indicates that an assembly of $N = 3$ cuboids with a defined shape anisotropy is already sufficient to benefit from dipolar interactions, with no further improvement by assembling more cuboids into the chain. Based on these results, we simplify the magnetosome chains by studying $N = 3$ cuboids arrangements, with the intrinsic advantage of reducing computational costs.

% According to this finding, we can describe the strain AMB-1 as more efficient, magnetically speaking, than MSR-1, as both strains benefit from dipolar coupling, but at less (metabolic) energy costs for AMB-1.

Figure~\ref{simulations}(\textbf{C}) includes the results of the hysteresis loops performed for the $N = 3$ chain for different Fe\textsubscript{2}O\textsubscript{3} $\delta$ values. As it can be seen, $H_{\mathrm{c}}$ is progressively reduced upon increasing the oxidation layer. Given that the anisotropy direction of the oxidized layer is random in every grain, together with the lower saturation magnetization of maghemite, a softer magnetic behavior of the MNP ensemble is expected. We have determined the influence of the intergrain coupling by testing two values for the exchange weakening, $k = 0.25$  and $0.75$, corresponding to low and high intergrain coupling scenarios, respectively. Although these values affect the absolute value of the coercivity, resulting in a lower coercivity for strongly-coupled grains, the trend followed by the decrease of the coercive field with increasing $\delta$ is similar in both cases (see Fig.~\ref{simulations}(\textbf{D})), indicating the crucial role of the deviations from the uniaxial anisotropy direction $\left< 111 \right>$ in the coercivity reduction.

Figure~\ref{simulations}(\textbf{D}) summarizes our micromagnetic results. We present the coercive fields as a function of $\delta$ for $N = 1$ and $3$ ($k = 0.25$ and 0.75), together with our experimental data. To facilitate the interpretation, we have normalized the $H_{\mathrm{c}}$ values to the coercivity at $\delta =$ 0, corresponding to $t = 0$. The error bars, below $5 \, \%$, reflect the average over 10 different random seeds per simulated angle. The decrease in coercivity as a function of $\delta$ is linear for both $N = 1$ and $3$ configurations, in agreement with the expected behavior of single-domain MNPs within the first-order approximation of the Stoner-Wohlfarth model~\cite{tannous2008stoner, de2013stoner, orue2018configuration}. The absolute value of the slope is smaller for $N = 3$ compared to $N = 1$ (by a factor of $\sim$0.85), which can be attributed to the enhanced net anisotropy term, that encloses both the intrinsic uniaxial anisotropy $K_\mathrm{{u}}$ and the dipolar interactions between the magnetosomes. This results reveals that, when oxidized MNPs are arranged in a chain, their average magnetization is similar to the one that individual less-oxidized MNPs. Therefore, preserving dipolar interactions among the MNPs would slow down the degradation of their magnetic properties over extended timescales compared to the isolated case.

Moreover, the intersection of our experimental data with the simulation results corresponding to $N =3$ and $k =0.75$ (dotted line in Figure~\ref{simulations}(\textbf{D}) allows to indirectly estimate the thickness of the oxidized layer in our samples. For MSR-1, we find oxidation percentages of approximately $25\%$ and $28\%$ after 3 and 6 years, respectively. In the case of AMB-1, the estimated oxidation levels are around $8\%$ and $16\%$ after 1.5 and 4 years, respectively. These values are unprecedented in comparison to existing literature, where oxidation rates exceeding $25\%$ are typically reported after only days or months~\cite{mickoleit2023long, kim2021slow, schwaminger2017oxidation}, highlighting the exceptional long-term stability of magnetite preserved in chains inside bacteria. 

We detect a significant slower degradation of AMB-1 compared to MSR-1, which we ascribe to the homogeneous size distribution of the former. In the Supplementary Information~\cite{SM_Nanoscale_2025}, we have analyzed the size distribution of both bacteria strains, finding a bimodal size distribution for MSR-1, centered at $\approx$ 15 and 43 nm. Whereas the larger MNPs have a reduced surface-to-volume ratio, reducing the impact of the oxidized shell, the smaller ones are more sensitive. As the macroscopic coercivity reflects an average of the individual contributions, oxidation effects are more present for the case of MSR-1 compared to AMB-1.

\subsection{Fingerprints of shell disorder in polarized SANS}
\label{SANS_pr}

The present study shows that the degradation of the magnetic properties of magnetite nanoparticles is compatible with the oxidation of the magnetite phase. This is accomplished by (i)~indirect experimental evidence, from the reduction of the Verwey transition and coercivity measured over the years, and by (ii)~micromagnetic simulations, which successfully reproduce this coercivity decrease when a magnetite@maghemite core@shell microstructure is modeled. While the combination of these approaches provides an estimate for the maghemite shell thickness $\delta$, only the direct experimental measurement of $\delta$ can validate such a hypothesis. M\"ossbauer spectroscopy is the experimental probe traditionally employed~\cite{channing1972study, schwaminger2017oxidation, gorski2010determination, almeida2015effect}. However, this technique does not provide enough resolution to measure MNPs in cell environments, as it is the case of the present study, where the magnetosomes are kept intra-cellular.

To detect and quantify thin oxide layers in MNPs embedded in realistic environments we propose here the use of polarized small-angle neutron scattering (SANS). This technique provides volume-averaged information with a spatial resolution that covers the relevant mesoscopic length scale (a few up to a few hundreds of nanometers)~\cite{muhlbauer2019magnetic, michels2021magnetic, zakutna2020field, gerina2024exploring, honecker2022using, gerina2024exploring}. Many examples in the literature demonstrate the success of magnetic SANS to resolve and quantify the magnetization distribution within iron oxide MNPs (e.g., \cite{disch2012quantitative, krycka2010core, krycka2014origin, zakutna2020field}). Furthermore, SANS has already been employed in magnetotactic bacteria to resolve their magnetic chain morphology~\cite{orue2018configuration} and the rotation mechanism under applied fields~\cite{bender2020probing}, validating this technique for the probe of MNPs embedded in cell environments.

%Furthermore, polarized SANS allows to access detailed information about the magnetic inhomogeneity in the sample, usually masked by the strong nuclear scattering, which can easily be removed when polarizing the neutron beam. More precisely, by analyzing the intensity in different spin states (spin-flip and non-spin-flip scattering), one can extract information about the magnetization distribution.

Based on the real-space micromagnetic simulation results, we calculate the SANS observables for a single magnetite@maghemite cuboidal MNP for different values $\delta$ of the maghemite surface-layer thickness. More specifically, we model the signatures of oxidation in the real-space pair-distance distribution function $p(r)$ of the spin-flip SANS cross section. This function describes the distribution of real-space distances between volume elements inside the particle, weighted by the excess scattering length density function~\cite{svergun2003small}. The spin-flip SANS cross section, from which we extract the $p(r)$ by Fourier transformation, can be measured in a uniaxial polarization analysis experiment (e.g., \cite{michels2010epjb,kryckaprl2014,zakutna2020,malyeyev2022,ukleev2024}) and does not contain the unwanted structural (nuclear coherent) scattering contribution that could potentially mask the fingerprints of magnetically disordered interfaces~\cite{michels2021magnetic}. Details regarding the micromagnetic SANS simulation procedure can be found in Refs.~\cite{michels2021magnetic, vivas2020toward, pratami2023micromagnetic, adams2024framework, sinaga2024neutron}.

\begin{figure*}
\centering
\resizebox{2.0\columnwidth}{!}{\includegraphics{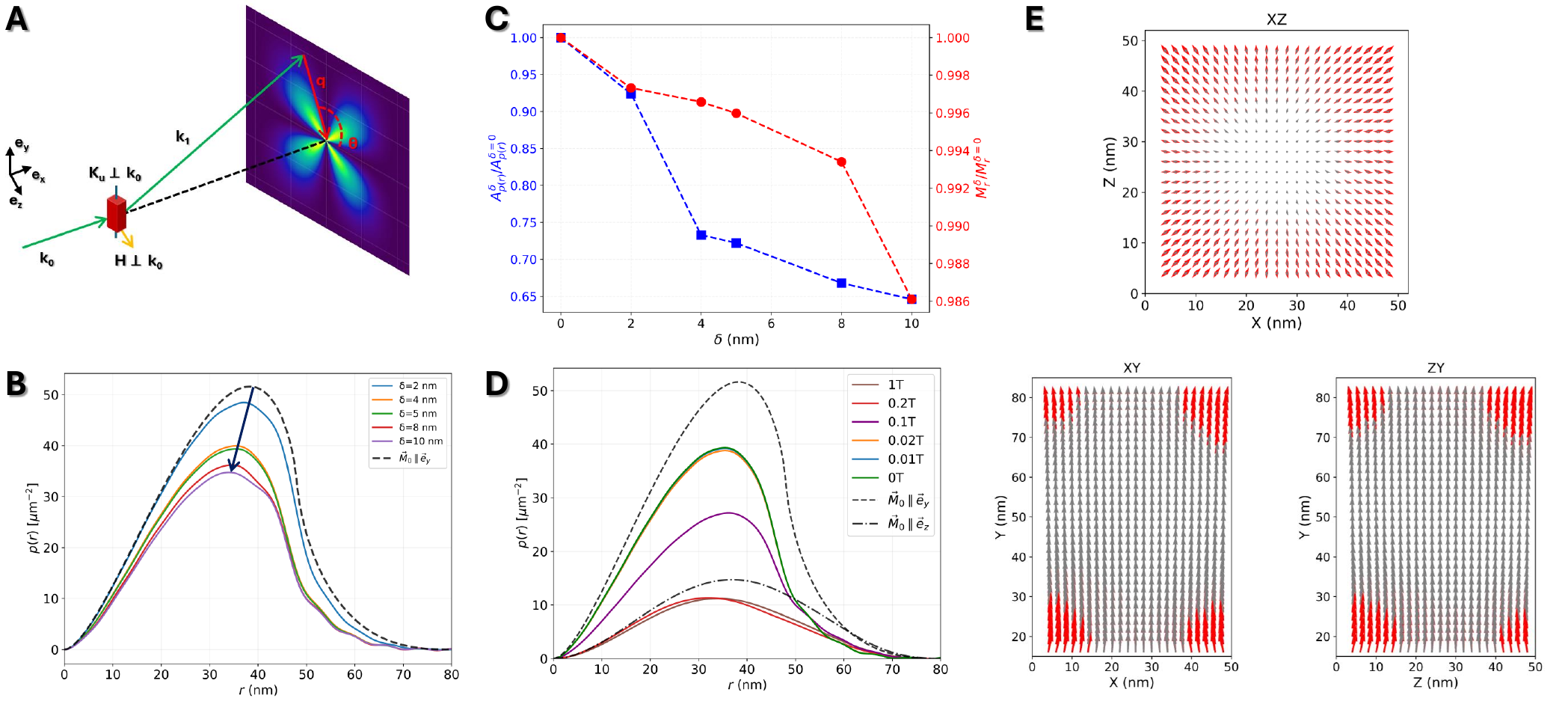}}
\caption{(\textbf{A})~Schematic representation of the scattering geometry: The uniaxial anisotropy axis of the particle $\vec{K}_{\mathrm{u}}$ is along the long axis of the cuboid, which is parallel (and antiparallel) to the $\vec{e}_y$ direction of a Cartesian laboratory coordinate system, and perpendicular to the wave vector $\vec{k}_0$ of the incident neutron beam ($\vec{k}_0 \parallel \vec{e}_x$). The externally applied magnetic field $\vec{H} \parallel \vec{e}_z$ is perpendicular to $\vec{K}_{\mathrm{u}}$. The particle has geometric dimensions of $48 \times 64 \times48 \;\mathrm{nm^{3}}$ and is oriented as shown. The momentum-transfer vector $\vec{q}$ is given by $\vec{q} = \vec{k}_0 - \vec{k}_1$. (\textbf{B})~Pair-distance distribution functions $p(r)$ at remanence for different shell thicknesses $\delta$ (see legend). The dashed line corresponds to the extreme case where the magnetization is completely saturated along $\vec{K}_{u} \parallel \vec{e}_{y}$, calculated using Eq.~(6) in the Supplementary Information~\cite{SM_Nanoscale_2025}. The arrow marks the displacement of the maximum towards smaller $r$ as $\delta$ increases. (\textbf{C})~Evolution of the ratio $A_{p(r)}^{\delta}/A_{p(r)}^{\delta = 0}$ (left axis) and of the normalized remanent magnetization $M_{\mathrm{r}}^{\delta}/M_{\mathrm{r}}^{\delta=0}$ (right axis) with the interface thickness $\delta$. (\textbf{D})~$p(r)$ for a series of applied fields (see legend) and for a fixed value of $\delta = 5 \, \mathrm{nm}$. The black lines represent the corresponding $p(r)$ of a cuboid uniformly magnetized along $\vec{e}_{z}$ (dotted-dashed line) and $\vec{e}_{y}$ (dashed line). These were, respectively, obtained using Eqs.~(5) and (6) in the Supplementary Information~\cite{SM_Nanoscale_2025}. (\textbf{E})~Remanent magnetization distributions shown for the three central planes of the cuboid ($\delta = 5 \, \mathrm{nm}$). The gray arrows represent magnetic moments that are oriented along the net particle magnetization, which is mainly along $\vec{e}_y$ with small components (below $2 \, \%$) along $\vec{e}_x$ and $\vec{e}_z$. The red arrows represent misaligned (canted) moments.}
\label{SANS}
\end{figure*}
%net particle magnetization: (0.01107018, 0.99983599, 0.01433329)

Figure~\ref{SANS}(\textbf{A}) depicts a sketch of the scattering geometry. In our calculations, we have considered only a single oriented magnetosome, as the surface oxidation effects are more pronounced at the level of individual particles than in the full magnetosome chain. All simulations are initialized from the saturated magnetic configuration. The easy-axis anisotropy direction $\vec{K}_{\mathrm{u}}$ is parallel (and antiparallel) to the $\vec{e}_y$ direction of a Cartesian laboratory coordinate system and perpendicular to the incoming neutron beam, which is parallel to $\vec{e}_x$. The externally applied magnetic field is along $\vec{e}_z$ ($\vec{H} \parallel \vec{e}_z$).

Figure~\ref{SANS}(\textbf{B}) features the computed $p(r)$ at remanence for the different $\delta$~values used in this work. We have included the results for a uniformly magnetized cuboid along its long axis as a reference for the ideal non-oxidized case (black dashed line). It is emphasized that the overall geometrical dimensions of the MNP ($48 \times 64 \times 48 \, \mathrm{nm}^3$) remain fixed in the simulations. The microstructure is defined as a magnetite core with dimensions of $(48-2\delta) \times (64-2\delta) \times (48-2\delta) \, \mathrm{nm}^3$, and a maghemite shell of variable thickness $\delta$ ($\delta = 0 - 10 \, \mathrm{nm}$); for instance, a cuboid with $\delta = 5 \, \mathrm{nm}$ consists of a $38 \times 54 \times 38 \, \mathrm{nm}^3$ magnetite core surrounded by a 5-nm thick maghemite shell. In all cases, the $p(r)$ in Fig.~\ref{SANS}(\textbf{B}) exhibit a slightly asymmetric profile, which is characterized by a linear-type rise at short distances and an extended smoother tail at larger $r$~values, in qualitative agreement with the result for a uniformly magnetized cuboid along its easy axis. This behavior underscores the dominant uniaxial anisotropy of the cuboid as the magnetization reorients from the field direction $\vec{e}_z$ at saturation to the anisotropy axis $\vec{e}_y$ at remanence. As $\delta$ increases, the maximum of $p(r)$ decreases and shifts to smaller $r$~values (as indicated by an arrow). The maximum of $p(r)$ shifts from $r_{\mathrm{max}} \cong 39 \, \mathrm{nm}$ for the uniformly magnetized case to $r_{\mathrm{max}} \cong 37 \, \mathrm{nm}$ for $\delta = 2 \, \mathrm{nm}$ and to the final value of $r_{\mathrm{max}} \cong 34 \, \mathrm{nm}$ for $\delta = 10 \, \mathrm{nm}$, indicating a reduction in the characteristic internal magnetic distances~\cite{adams2022magnetic_2, adams2022magnetic}. This observation is compatible with an increasing degree of magnetic disorder (when $\delta$ increases), as the surface layer is composed of cells with random anisotropy directions.

We propose to use the area under the $p(r)$~curve to extract quantitative information on the degree of magnetic disorder as a function of $\delta$. It is well known from the theory of magnetic SANS that the value of the azimuthally-averaged spin-flip scattering intensity at the origin of reciprocal space, $I_{\mathrm{sf}}(q=0)$, and the integral of the corresponding pair-distance distribution function are related via~\cite{svergun2003small}
\begin{eqnarray}
I_{\mathrm{sf}}(q=0) = \int_0^{\infty} p(r) dr = A_{p(r)} .
\label{eq-pr}
\end{eqnarray}
By taking the value at $\delta = 0 \, \mathrm{nm}$ as a reference, we can consider the following ratio
\begin{eqnarray}
\frac{I_{\mathrm{sf}}^{\delta}(q = 0)}{I_{\mathrm{sf}}^{\delta = 0}(q = 0)} = \frac{A_{p(r)}^{\delta}}{A_{p(r)}^{\delta = 0}} ,
\label{eq-ratio}
\end{eqnarray}
which is compared to the normalized remanent magnetization $M_{\mathrm{r}}^{\delta}/M_{\mathrm{r}}^{\delta=0}$ obtained from the micromagnetic simulations. At this point, we would like to recall a crucial difference between magnetization and SANS data: while the magnetization in a conventional magnetometry experiment measures the spatially-averaged net magnetic moment of the entire particle ensemble, SANS provides Fourier-space-resolved mesoscale information of spatial correlations of the local magnetization. As such, both techniques are complementary but not directly equivalent. Figure~\ref{SANS}(\textbf{C}) compares the evolution of the above-defined quantities as a function of $\delta$. As expected, both decrease with increasing $\delta$, but their trends differ markedly. Whereas the magnetization data change very weakly, by only $\cong 1.5 \, \%$ for the largest $\delta = 10 \, \mathrm{nm}$, the neutron results vary more strongly, by more than $30 \, \%$ for the same $\delta$ variation. These findings underscore the superior sensitivity of the SANS technique to nanoscale magnetic spin disorder. Motivated by this, and given that the quantity $I_{\mathrm{sf}}^{\delta}(q = 0)$ is related to the average magnetization (as it is accessible in a magnetometer), future studies should aim to {\it quantitatively} relate the ratio Eq.~(\ref{eq-ratio}) to $\delta$.

Figure~\ref{SANS}(\textbf{D}) displays the field dependence of $p(r)$ for an interface thickness of $\delta = 5 \, \mathrm{nm}$, together with the reference curves corresponding to the uniformly magnetized cuboid along $\vec{e}_z$ and $\vec{e}_y$ (calculated from Eqs.~(5) and (6) in the Supplementary Information~\cite{SM_Nanoscale_2025}). At the highest field ($\mu_{0} H = 1 \, \mathrm{T}$), the magnetization is (nearly) uniformly aligned along the $\vec{e}_z$~direction (a short axis of the cuboid), giving rise to a nearly symmetric bell-shaped $p(r)$~distribution that very closely resembles the $p(r)$ of the uniformly magnetized cuboid along $\vec{e}_{z}$ (black dotted-dashed line). The maximum of the $p(r)$ for $\delta = 5~\mathrm{nm}$ is slightly shifted to smaller $r$ and slightly below the one of the uniformly magnetized cuboid, indicating the existence of minor spin misalignment event at $1 \, \mathrm{T}$, which is not detectable in the magnetization measurements (compare to Fig.~\ref{simulations}(B)). As the external field is reduced, the magnetization direction reorients from $\vec{e}_z$ towards the magnetic easy axis $\vec{K}_{\mathrm{u}} \parallel \vec{e}_y$, which coincides with the long axis of the cuboid. This rotation results in a progressive asymmetry of $p(r)$ and in an increase of its overall magnitude, while the position of the maximum shifts slightly to larger $r$ values ($r_{\mathrm{max}}^{1 \, \mathrm{T}} \cong 34.0 \, \mathrm{nm}$ and $r_{\mathrm{max}}^{0 \, \mathrm{T}} \cong 35.5 \, \mathrm{nm}$) as the magnetization reorients from $\vec{e}_z$ to $\vec{e}_y$ with decreasing field. Both effects signal the development of a magnetically disordered shell at low fields, which results in a significant reduction of the characteristic magnetic distances~\cite{adams2022magnetic_2, adams2022magnetic}.

To further visualize our results, we show in Fig.~\ref{SANS}(\textbf{E}) planar cuts through the 3D remanent magnetization distribution for a fixed value of $\delta = 5 \, \mathrm{nm}$. The average magnetization of the cuboid is depicted by gray arrows, while the red arrows represent spins that deviate by more than $2 \, \%$ from the mean direction. As is seen, the magnetic disorder is primarily localized at the particle edges, where the stray field is large and induces strong torques on the magnetic moments. The spin deviations remain modest, typically less than $5 \, \%$. In other words, the magnetization distribution $\vec{M}(\vec{r})$ within the core part of the cuboid is relatively uniform and the largest gradients in $\vec{M}(\vec{r})$ appear at the interface region, which represents a perturbation in the magnetic microstructure (different magnetic materials parameters and a random distribution of anisotropy axes). The spin orientation varies smoothly across the interface region and $\vec{M}(\vec{r})$ within the shell region is not randomly oriented, as the assumed anisotropy distribution might impose, but it smoothly deviates from the core direction (by less than $5 \, \%$). These results suggest that polarized SANS, unlike bulk magnetometry, provides direct access to internal microstructural features that govern the magnetic behavior at the nanoscale, features that remain effectively invisible to macroscopic magnetization measurements.

%%%%%%%%%%%%%%%%%%%%%%%%%%%%%%%%%%%%%%%%%%%%%%%%%%%%%%%%%%%%

\section{Summary and Conclusion}
\label{conclusions}

In this work, we have provided a micromagnetic description for the oxidation-driven transformation of magnetite to maghemite in magnetite nanoparticles. This approach captures the interplay between magnetic order and disorder at the interface and resolves the magnetization of the nanoparticles with a nanometer resolution, going beyond the conventional single-domain picture. Our simulations incorporate dipolar interactions to capture realistic interparticle interaction effects and are benchmarked against experimental data from the magnetosome chains grown by magnetotactic bacteria. By keeping the magnetosomes within the bacteria body, the magnetic properties are preserved over an unprecedented time scale of several years. This has been cross-checked by following the oxidation process for two different bacterium strains.

Our experimental results in magnetotactic bacteria show a remarkable slower oxidation compared to the one in magnetite nanoparticles reported in the literature (years vs. days). This fact, together with our micromagnetic results, which indicate a slower decrease of coercivity with $\delta$ at chain configurations, allows us to conclude on the beneficial role of dipolar interactions, and the related shape anisotropy, to keep the hysteresis loop broad compared to the isolated MNP situation. As a consequence, equally oxidized MNP will preserve their magnetic properties longer when they are arranged in chains, compared to their individual-particle response. We therefore propose a strategy to preserve the magnetic properties of magnetite at longer time scales by encapsulating them and letting them interact, mimicking the natural protection of the bacterial body. A very recent example for such a strategy includes the biomineralization of iron oxide in encapsulin nanocompartments by Efremova~\textit{et al.}~\cite{efremova2025genetically}.

Furthermore, for the example of a single oriented cuboid, we have resolved the spin distribution at the interface between magnetite and maghemite phases by numerically calculating the pair-distance distribution function $p(r)$ extracted from the polarized small-angle neutron scattering (SANS) cross section. We have connected the increase of the magnetic disorder in the shell region to the reduction of the remanent magnetization via the area under the $p(r)$~curve. Our results are compatible with a smoothly varying magnetization distribution across the shell (magnetically diffuse interface), which contrasts with the geometrically sharp interface. %Further studies should combine experiments and calculations to quantitatively relate $\frac{A_{p(r)}^{\delta}}{A_{p(r)}^{\delta = 0}}$ to $\delta$ in absolute units, resolving the precise contribution of each Fourier component of the magnetization to the spin-flip SANS cross-section. 

The disparity between geometric and magnetic dimensions might account for the prolonged retention of magnetic functionality in partially oxidized magnetite MNPs. Our results challenge the assumption that structural oxidation boundaries directly dictate magnetic degradation, and they highlight the power of integrating micromagnetic simulations with polarized SANS to reveal hidden aspects of magnetic (dis)order in complex nanoscaled systems. Beyond its relevance for fundamental studies of disorder, this approach opens the door to characterizing magnetic nanoparticles embedded in cellular environments (which are compatible with SANS), where their performance differs significantly from that observed in extra-cellular experiments~\cite{calaf2018oxidative, parolini2009microenvironmental, gandarias2023intracellular}.

%%%%%%%%%%%%%% Author contribution statement

\section*{Author contributions}

Conceptualization: E.M.J.
Methodology: E.M.J.
Software: E.M.J., M.A., J.L.
Validation: E.M.J., M.A., A. M., J.L.
Formal analysis: E.M.J.
Investigation: E.M.J., J.A.
Sample preparation: L.G., L.M.
Resources: E.M.J., J.A., A.M., J.L.
Data curation: E.M.J., J.A.
Visualization: E.M.J.
Supervision: A.M., J.L.
Project administration: E.M.J., A.M.
Funding acquisition: E.M.J., A.M.
Writing original draft: E.M.J.
Writing, review, and editing: All authors.

%%%%%%%%%%%%% Data

\section*{Data availability}

All data supporting the findings of this study are included within the article and any ESI.

%%%%%%%% Conflicts of interest

\section*{Conflicts of interest}

There are no conflicts to declare.

%%%%%%%%%% Acknowledgements

\section*{Acknowledgements}

E.M.J. acknowledges funding from the European Union's Horizon 2020 research and innovation program under the Marie Sk{\l}odowska-Curie actions grant agreement No.~101081455-YIA (HYPUSH); the Institute for Advanced Studies (IAS) of the University of Luxembourg; and the Fonds Wetenschappelijk Onderzoek (FWO-Vlaanderen) project number No.~V501325N. M.P.A. and A.M. gratefully acknowledge the financial support from the National Research Fund of Luxembourg (AFR Grant No.~15639149). L.G. acknowledges funding from the European Union’s Horizon 2020 research and innovation program under the Marie Sk{\l}odowska-Curie grant agreement No.~101150206 (DroneMTB). L.M. thanks the Horizon Europe Programme for the financial support provided through a Marie Sk{\l}odowska-Curie grant agreement No.~101067742 (ProteNano-MAG). J.L. is supported by the FWO-Vlaanderen with senior postdoctoral research fellowship No.~12W7622N. The authors acknowledge SGIker (UPV/EHU/ERDF, EU) for technical and human support, and resources and services provided by the HPC facilities of the University of Luxembourg. 

%%%%%%%%%%%%%%%  References  %%%%%%%%%%%%%%%%%%%%%%%

\bibliography{references}

\end{document}